\documentclass[conference]{IEEEtran}
\usepackage{amssymb}
\usepackage[cmex10]{amsmath} 
\usepackage{cite} 
\usepackage{url} 
\usepackage{url}
\usepackage{color,soul}
\usepackage{ifpdf}
\ifCLASSINFOpdf
  \usepackage[pdftex]{graphicx}
  \DeclareGraphicsExtensions{.pdf}
\else
  \usepackage[dvips]{graphicx}
   \DeclareGraphicsExtensions{.eps}
\fi
\usepackage{pgfplots}

\begin{document}
\title{Towards a Cognitive Routing Engine for Software Defined Networks}
\author{Frederic Francois and Erol Gelenbe\\
Intelligent Systems and Networks\\
Department of Electrical and Electronic Engineering\\
Imperial College, London SW7 2BT, UK\\
Email: \{f.francois and e.gelenbe\}@imperial.ac.uk\\
}
\maketitle

\begin{abstract}
Most Software Defined Networks (SDN) traffic engineering applications use excessive and frequent global monitoring in order to find the optimal Quality-of-Service (QoS) paths for the current state of the network. In this work, we present the motivations, architecture and initial evaluation of a SDN application called \textit{Cognitive Routing Engine} (CRE) which is able to find near-optimal paths for a user-specified QoS while using a very small monitoring overhead compared to global monitoring which is required to guarantee that optimal paths are found. Smaller monitoring overheads bring the advantage of smaller response time for the SDN controllers and switches. The initial evaluation of CRE on a SDN representation of the GEANT academic network shows that it is possible to find near-optimal paths with a small optimality gap of 1.65\% while using 9.5 times less monitoring.  
\end{abstract}

\section{Introduction}
The ossification of computer networks due to the use of non-standardized and closed source protocols between the control and data plane of commercial Network Forwarding Elements (NFEs) has led researchers to develop Software Defined Networks (SDNs) \cite{feamster2013} where open and standardized protocols, such as OpenFlow (OF) \cite{of_1_3_5}, are used to program the data plane of NFEs. In order to route traffic, SDNs usually make use of routing applications which runs on top of the SDN controller. These routing applications make routing decisions based on network policies and the state of the network but unfortunately, gathering the state of the network is an expensive activity both in terms of processing overhead at the controller and NFEs and control traffic \cite{yassine2015}. With this in mind, we develop a new routing application called \textit{Cognitive Routing Engine} (CRE) which significantly increases the efficiency of the network state gathering process while obtaining enough information about the network to calculate the best paths that meet the Quality of Service (QoS) requirements of the host applications which use them. 
The design objectives of CRE are:
\begin{itemize}
\item the state of the network should be gathered in an efficient way, i.e. a minimum number of OF messages should be exchanged between the SDN controller and the NFEs during the gathering process since each OF message has a processing overhead and occupies bandwidth on the control network of the SDN;
\item the optimization of the network monitoring process should not lead to worse paths being found by the CRE SDN application;
\item the use of CRE should not lead to frequent changes in the path chosen by CRE for a given traffic flow, i.e. no frequent oscillations of the path for a given flow by CRE which can increase the control plane processing and bandwidth overhead inside SDNs;
\item updates to the flow tables of the NFEs must not have an adverse impact on the traffic being transported by the NFEs. For e.g., no packet loss and forwarding loops when flow tables are updated; and
\item the implementation of CRE should be compatible with existing SDN-enabled NFEs so as to enable more network operators to adopt CRE without purchasing new hardware and software, and avoid lengthy standardization processes. 
\end{itemize}

\section{Related Work}
In current literature, most traffic engineering techniques \cite{akyildiz2014} used in SDN takes the view that SDN provides a global view of the current network state and topology in a logically centralized controller and therefore, optimization can be performed on the network by running different types of global traffic engineering algorithms such as constrained shortest path first. In this paper, we take the view that networks can be large and therefore, it is inefficient to obtain accurate and updated state information for the whole network at a frequency required for effective traffic optimization. This view is supported by the large amount of work that is currently undertaken to reduce the excessive overhead linked with accurate and frequent global monitoring the network \cite{yassine2015}.    
Existing work on network monitoring in SDN can be divided into 2 categories: active and passive monitoring \cite{yassine2015}. Active monitoring involves either the actual probing of the network by sending special packets \cite{phemius2013} and/or polling the state of the network through OF mechanism, e.g. retrieving the value of OF counters \cite{of_1_3_5}, while passive monitoring either only observes the existing SDN behaviour to infer the state of the network, e.g. \textit{Packet-In} and \textit{Flow Removed} OF control messages \cite{yu2013}, or  calculates the  network state based on the collected sampled packets \cite{suh2014}.

The Cognitive Routing Engine developed in this paper is similar to the routing algorithm that is used in Cognitive Packet Networks (CPNs)\cite{gelenbe1999,gelenbe2000,gelenbe2001,gelenbe2002} but with the significant and important differences that in CPN, each router runs its own routing and learning algorithm for every source-destination pair and QoS requirement and, the routers collect the network state through the use of smart packets and their associated acknowledgment packets. This is in contrast with CRE which runs in a logically centralized manner on top of the SDN controller and uses exclusively OF mechanisms to gather network state. Hence, CRE can be viewed as a SDN-compatible version of CPN and can be used with the large number of SDN-enabled switches that has already been deployed. CPN has been used successful in various settings \cite{sakellari2010}, for e.g.,  traffic engineering\cite{gelenbe2003}, routing in wireless \cite{gelenbe2004} and sensor \cite{hey2008} networks and defence against Denial-of-Service (DoS) attacks \cite{gelenbe2005}.

The main contribution of this paper is a SDN-compatible CPN application which reduces the amount of network monitoring while maintaining the ability to discover paths which meet the QoS requirements of host applications.




\section{Overall Archictecture}

\begin{figure}[]\centering
\includegraphics[width=0.49\textwidth]{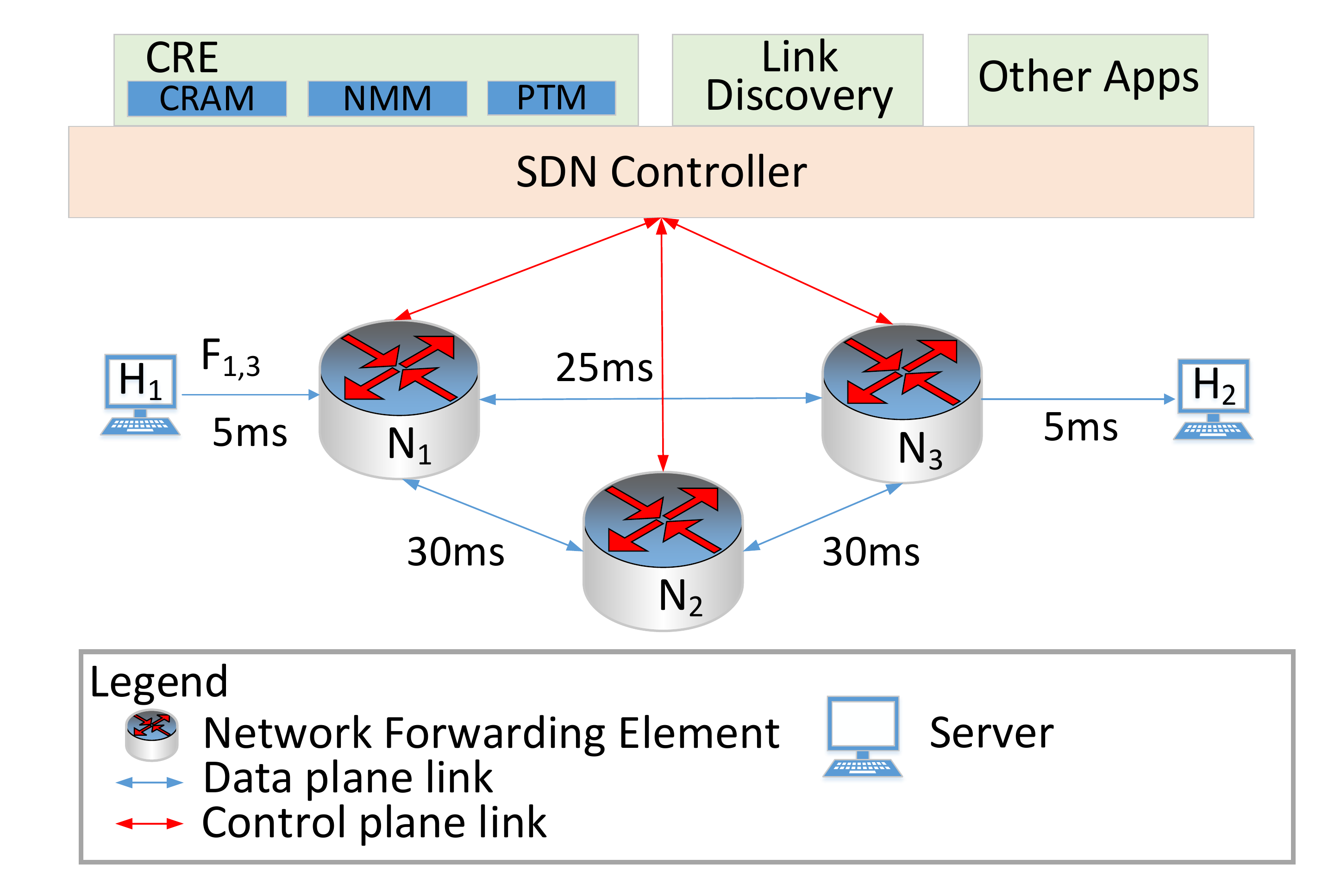}
\caption{Overall network architecture showing where the Cognitive Routing Engine is located and how it interacts with other components of the architecture.}
\label{fg:net_arch}
\end{figure}

Fig.~\ref{fg:net_arch} shows where the CRE application is located in the overall SDN architecture and how it interacts with the various other components of a typical SDN deployment.

A brief description of each component of the overall SDN architecture is provided as follows:

\textbf{Network Forwarding Elements} (NFE)---are SDN-enabled packet switches and use OpenFlow v1.3~\cite{of_1_3_5} as the communication protocol between the control and data plane of the NFEs in this particular instance.

\textbf{SDN Controller}---a.k.a Network Operating System, is responsible to send and receive OF messages from the NFEs. In addition, the SDN controller typically parses the received OF messages into data structures which can be understood by the different applications running on top of the controller.

\textbf{Applications}---are pieces of software running on top of the controller which provides specialized network functions such as Network Address Transation (NAT) and firewall. In this particular instance, there are 2 applications:
\begin{itemize}
\item Link Discovery application---which discovers the data plane topology of the network by using the standardized Link Layer Discovery Protocol (LLDP) \cite{lldp_std_2009}. A link is defined in this paper as unidirectional. 
\item CRE application---which efficiently finds and installs new suitable paths in the network as requested by other SDN applications. 
CRE is made of 3 main modules: 
\begin{itemize}
\item the Cognitive Routing Algorithm Module (CRAM)---which uses Random Neural Networks (RNNs) with Reinforcement Learning (RL) to find network paths which maximize a customizable objective function and therefore, meet the QoS requirements of host applications,
\item a Network Monitoring Module (NMM)---which efficiently either uses past network measurements and/or probes and/or poll the network to get the necessary network state information to update the RNNs in the CRAM.
\item a Path-to-OF Translator Module (PTM)---which is able to convert the paths found by the CRAM into the appropriate set of OF messages so that paths are either created or updated with minimum inconsistency in the network. 
\end{itemize}
\end{itemize}

\section{Message Exchange Sequence between different Components of SDN Architecture}

\begin{figure}[]\centering
\includegraphics[width=0.49\textwidth]{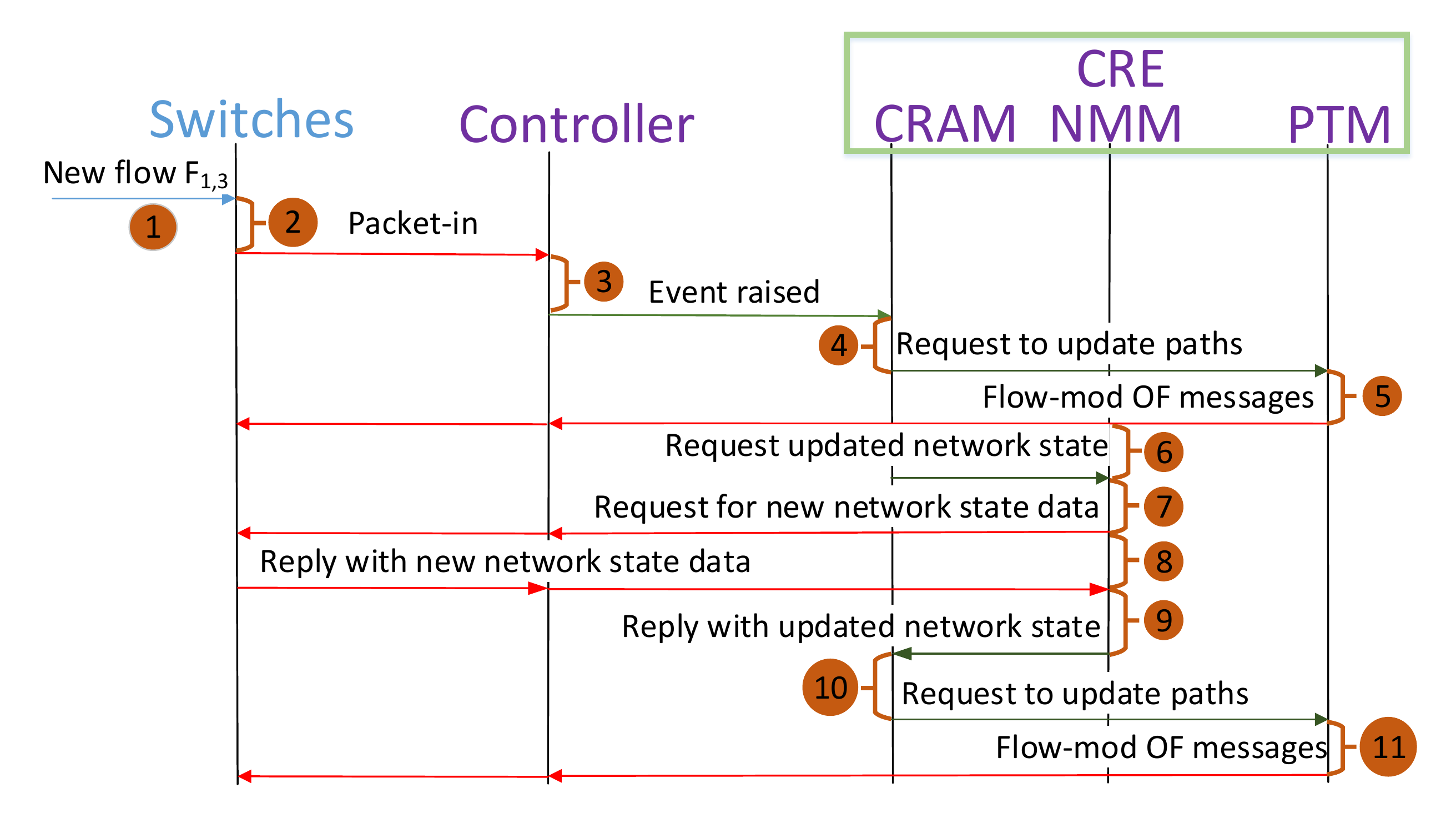}
\caption{Message exchange sequence between different components of the network during an initial flow set-up.}
\label{fg:msg_seq_1}
\end{figure}

Fig.~\ref{fg:msg_seq_1} provides an example of the sequence of message exchange between the different components of the SDN architecture when a new flow arrives at the network and paths need to be set up to route the flow through the SDN according to the network policies set by the network operator. A flow in the electrical domain is considered by OF as a collection of packets where the  value of a subset of the layer 2 to 4 fields are the same and a new flow is one where there is no rule which matches the flow inside the flow table(s) of the first encountered NFE.

The different steps in Fig.~\ref{fg:msg_seq_1} are described below:
\begin{itemize}
\item Step 1: A new flow $F_{1,3}$ arrives at the NFE $N_1$ which needs to travel to NFE $N_3$.
\item Step 2: NFE $N_1$ analyzes the first packet of the flow $F_{1,3}$ and finds that there is no rule inside its flow table(s) which matches the packet. Therefore, $N_1$ encapsulates the packet into an OF \textit{Packet-In} message and sends the message to its master SDN controller since the default action for a \textit{table-miss} event was set to forward the packet to the master controller during the configuration of the NFE.
\item Step 3: The master SDN controller receives the OF \textit{Packet-In} message and parses it into a data structure more suitable for processing by different applications running on top of the SDN controller. The SDN controller notifies and sends the data structure to all SDN applications which have registered with it to receive the OF \textit{Packet-In} messages. 

%

\item Step 4: Notified applications decide if they are interested to perform actions based on the data structure that is sent by the controller. In this particular scenario, the CRE application is the only relevant application which will perform actions due to the new flow $F_{1,3}$.

\item Step 5 \& 6: CRE first installs a path for flow $F_{1,3}$ by calculating the shortest path based on hop count between NFE $N_1$ and $N_2$ by using the network topology discovered by the Link Discovery application. This allows packets to be routed in the network fast without waiting for CRE to collect network measurements and find a path based on RNN which can lead to packet loss.  The PTM module of CRE is responsible to implement the new path for flow $F_{1,3}$ in the NFEs by installing the appropriate OF rules as will be described in Section \ref{lb:PTM}.

Next, CRAM finds the most suitable links for a given path request by using RNNs with RL. In order for CRAM to operate, it needs to gather information about the network state, which is the responsibility of the NMM in CRE. The detailed description of how CRAM operates will be given in the following Section \ref{lb:CRAM}.

\item Step 6, 7 \& 8: The NMM receives requests from the CRAM for certain characteristics of links in the network so that CRAM can calculate the numeric value of the objective function of the whole path that it has chosen and also updates its RNNs so that it chooses better paths in the future.
Currently, NMM can obtain utilization, quality (based on packet loss and frame errors) and delay of the links of the network. NMM can retrieve and calculate link utilization and quality of a given link by using the OF \textit{Port Stats Request} messages. On the other hand, OF doesn't have any native ability to obtain the necessary information from the network in order to calculate the packet delay on a link but the SDN controller can send probe packets to obtain a good estimate of packet delays \cite{phemius2013}. In addition, NMM is intelligent enough that it does not always query the network for new data but it can use its database of recently gathered data to reply to the requests made by CRAM. This increases the efficiency of the network monitoring process for the whole CRE.

\item Step 9: CRAM updates its RNNs through reinforcement learning based on the data received from the NMM and will output the best paths found. If paths need to be updated, they are then passed to the PTM.

\item Step 10: The PTM is responsible to implement the paths in the NFEs by modifying the flow table(s) of the NFEs in a way which will not lead to packet loss and forwarding loops in the network. This is explained in detail in Section \ref{lb:PTM}.
\end{itemize}

\section{CRAM: Cognitive Routing Algorithm Module}
\label{lb:CRAM}
CRAM is a routing algorithm which is based on Random Neural Network (RNN) with Reinforcement Learning (RL) \cite{gelenbe1999,gelenbe2000,gelenbe2001,gelenbe2002}. For a given flow, a new RNN is created for each NFE along the path initially found through shortest path routing. 
A RNN consists of a network of interconnected neurons where each neuron represents one active port of the NFE. The decision of the RNN on which output to use as the next hop can be either exploratory or exploitation. When the RNN is in exploratory mode, it chooses the output port randomly while when the RNN is in exploitation mode, it chooses the neuron which has the highest potential as will be described in the next paragraph. Exploratory mode is chosen with a probablity $X$\%, e.g. 5\%\cite{sakellari2010}, of the time and exploitation mode is chosen for the remaining $(1-X)$\% of the time. If the next hop NFE does not currently have a representative RNN for this particular flow, a new RNN is created for it.

Each neuron has a property called potential and the neuron with the highest potential is selected as the output when the RNN is in exploitation mode and therefore, the most excited neuron determines the next hop NFE of the path. The potential, $q_i$, of neuron $i$ can be calculated by using:
\begin{equation}
\label{eq:rnn_q}
q_i=\frac{\lambda^{+}_i}{r_i + \lambda^{-}_i}
\end{equation}
where $\lambda^{+}_i$ and $\lambda^{-}_i$ are respectively the total positive and negative potential that are transferred from other neurons connected to neuron $i$ in the RNN. These total potentials are dependent on both the potential of the neurons themselves and the weights of the links and can be calculated as:
\begin{align}
\label{eq:rnn_lambda}
\lambda^{+}_i=\sum\limits_{j \in N} q_j w^{+}_{j,i} + \Lambda_i^{+} ~~~~\text{where}~j \neq i\\\nonumber
\lambda^{-}_i=\sum\limits_{j \in N} q_j w^{-}_{j,i} + \Lambda_i^{-} ~~~~\text{where}~j \neq i
\end{align}
where $w_{j,i}^+$ and $w_{j,i}^-$ are respectively the positive and negative weight of the link from the other neurons $j$ in the RNN to the neuron $i$ and $N$ is the set of all neurons in the RNN. $\Lambda_i^{+}$ and $\Lambda_i^{-}$ are the fixed rate at which external positive and negative spikes arrives at the neuron respectively.  $r_i$ is a normalization factor (firing rate in \cite{sakellari2010}) which is calculated as:
\begin{equation}
\label{eq:rnn_r}
r_i=\sum\limits_{j \in N} [w^{+}_{i,j} + w^{-}_{i,j}] ~~~\text{where}~ j \neq i
\end{equation}

The positive $w^{+}_{i,j}$ and negative $w^{-}_{i,j}$ weight of each link are determined with the help of reinforcement learning. The current exponential averaged value of the objective function $O_{sd}[t]$ of a path with source $s$ and destination $d$ is calculated as follows:
\begin{equation}
\label{eq:rnn_objective}
O_{sd}[t] = \alpha O_{sd}[t-1] + (1- \alpha)(o_{sd}[t])~~~\text{where}~0 \textless \alpha \textless 1
\end{equation}
where $O_{sd}[t-1]$ is the previous value of the objective function, $o_{sd}[t]$ is the current value of the objective function and $\alpha$ is the exponential average parameter which determines the relative importance of the current value of the objective function compared to the previous values. If the objective function is to be minimized, the reward $R_{sd}[t]$ of the path is calculated as $R_{sd}[t]=\big(O_{sd}[t]\big)^{-1}$. The effect of $R_{sd}[t]$ on the link weights in all RNNs along the found path depends on whether $R_{sd}[t]$ is greater or less than a threshold $\gamma_{sd}[t]$ with $\gamma_{sd}[t]$ calculated as follows:
\begin{equation}
\label{eq:rnn_objective}
\gamma_{sd}[t] = \beta \gamma_{sd}[t-1] + (1- \beta)R_{sd}[t]~~~\text{where}~0 \textless \beta \textless 1
\end{equation}
where $\beta$ is the exponential average parameter.

If $R_{sd}[t] \geq \gamma_{sd}[t]$, i.e. the reward of the new path is greater than the threshold and therefore the RNNs have made the right decision, the link weights in each RNN is updated as follows: $w^{+}_{j,i}=w^{+}_{j,i} + R_{sd}[t]$ and $w^{-}_{j,k}=w^{-}_{j,k} + \frac{R_{sd}[t]}{|N|-2}$ where $j \in N$ and $k \neq i$. In the first link weight update operation, the positive link weights from other neurons $j$ to the chosen neuron $i$ are increased so that the potential $q_i$ of neuron $i$ is increased and the port associated with neuron $i$ is chosen again as output for the next search if the RNN is in exploitation mode. In the second link weight update operation, the negative link weights from neurons $j$ in the RNN to other neurons other than $i$ are then increased so that their potential is decreased with as end result the reduction of the probability that their associated ports will be chosen as output if the RNN is in exploitation mode.

On the other hand, if $R_{sd}[t] \leq \gamma_{sd}[t]$, i.e. the reward of the new path is less than the threshold and therefore the RNNs have made the wrong decision, the link weights in each RNN is updated as follows: $w^{+}_{j,i}=w^{+}_{j,i} + \frac{R_{sd}[t]}{|N|-2}$ and $w^{-}_{j,k}=w^{-}_{j,k} + R_{sd}[t]$ where $k \neq i$. In the first case, the positive link weights from neurons $j$ in the RNN to other neurons other than $i$ are increased so that their potential is increased and it is more likely that one of them will have its associated port chosen as the output during the next search rather than the previously chosen associated port of neuron $i$. In the second case, the negative link weights from other neurons $j$ to the chosen neuron $i$ are increased so that the potential $q_i$ of neuron $i$ is decreased with as end result the reduction of the probability that the port associated with $i$ will be chosen as output during the next search.

In order to prevent the link weights $W$ to increase indefinitely, they are re-normalized by first calculating the new $r_i^*$ by using the updated $W$ values in eq.~\eqref{eq:rnn_r} and updating the link weights as follows: $W=W\frac{r_i}{r_i^*}$.

Furthermore, CRAM keeps a record of the best $Z$ recent paths found for each path request and the best path out of the $Z$ paths is used for the actual routing of the traffic flow if the existing path has been active for longer than minimum time period to avoid frequent path fluctuations. 

\section{NMM: Network Monitoring Module}
\label{lb:NMM}
The Network Monitoring Module (NMM) is responsible for providing to the CRAM and external SDN applications the characteristics of the links that they request. The link characteristics that can currently be monitored and calculated by NMM are: utilization, quality (based on packet loss and frame errors) and delay. It should be noted that NMM cannot directly measure these characteristics but can obtain them after performing operations on measurement information provided by the NFEs and sending probe packets if necessary.   
The NMM is made of the following components:
\begin{itemize}
\item \textbf{Northbound API}---allows the NMM to interact with the CRAM and external SDN applications. The characteristics of links can be requested to the NMM and the NMM can also receive information from other SDN applications, for e.g., the links which are active in the network can be received from the Link Discovery application \cite{lldp_std_2009}.
\item \textbf{NFEs Database}---stores information about the fixed characteristics of NFEs. During the initial contact between the NFEs and the SDN controller, the NFEs send information about their identity and the features that they support by sending OF \textit{Features Reply} messages \cite{of_1_3_5} (it should be noted that the port description in OF v1.3 \cite{of_1_3_5} is now obtained through the use of OF \textit{Port Description Request/Reply} messages). The following information are stored: NFE Data Plane Identifier (DPID), and the port number, MAC address and speed (as negotiated between 2 adjacent NFEs) of the different ports of the NFEs. 
\item \textbf{Links Database}---stores information about the characteristics of each unidirectional link in the SDN. There are 2 types of characteristics stored about the links: fixed and non-fixed. Fixed characteristics of the links are: source NFE DPID $s$, destination NFE DPID $d$, source port number at $s$, destination port number at $d$ and capacity $b$. The non-fixed characteristics are: operational status, utilization, quality and delay. In addition, each non-fixed characteristic of the link has an associated timestamp field which stores the time when the characteristic was last updated.
\item \textbf{Link Delay Monitoring}---sends and receives probe packets in the network in order to measure the delay of a link in a similar way as in \cite{iyer2013}. The detailed procedure will be described below.
\item \textbf{OF Statistics Gathering}---send (resp. receive) OF \textit{Port Stats Requests}  (resp. \textit{Replies}) to calculate the utilization and quality of a link. The detailed procedure will be described below.
\item \textbf{Southbound API}---allows the NMM to interact with the SDN controller in order to send and receive OF statistics messages as well as probing packets for delay measurements. 
\end{itemize}

\subsection{Link Delay Monitoring}
A Link Delay Monitoring mechanism based on using only the features that OF \cite{of_1_3_5} provides so that CRE is compatible with OF-enabled NFEs which are already deployed in the field. Since the OF protocol as of \cite{of_1_3_5} does not specify that an OF-compatible NFE can measure and store link delay by itself, it is necessary to make use of probe packets in order to measure and calculate the delay of the link between 2 NFEs. The Link Delay Monitoring mechanism follows the following steps to calculate the link delay between 2 NFEs $N_s$ and $N_d$ with source and destination port $P_s$ and $P_d$ respectively:
\begin{itemize}
\item Step 1: An OF \textit{Echo Request} message (with a specified transaction id \textit{xid}) along with an OF \textit{Barrier Request} message is sent to NFE $N_s$ and the time $T_{N_s}$ it takes for $N_s$ to reply back with a corresponding OF \textit{Echo Reply} (identified by the \textit{xid} value) is measured. The same is done for $N_d$.

\item Step 2: The SDN controller is instructed to send an OF \textit{Packet Out} message to NFE $N_s$ where the source and destination MAC of the packet is set to the MAC address of the port $P_s$  and port $P_d$ respectively. The MAC addresses are obtained by using an OF \textit{Port Description Request} message in OF v1.3\cite{of_1_3_5} during the population of the NFEs Database. The \textit{Packet Out} message also contains the single action of outputing the packet on the port $P_s$. The Ethernet protocol type of the packet is set to an arbitrary value, e.g. 0x07c3, so that the SDN controller can identify the packet as a probe packet. The time $T_t$ that it takes for the SDN controller to receive the probe packet from $N_d$ is measured. The delay $D_{sd}$ between $N_s$ and $N_d$ is calculated as $D_{sd}=T_t-\frac{T_{N_s}}{2}-\frac{T_{N_d}}{2}$

\item Step 3: The NMM will instruct the SDN controller to periodically send probe packets as in step 2 above if the link continues to be selected by any RNN for probing. $T_{N_s}$ and $T_{N_d}$ are periodically measured also since these may change during the network operation.
\end{itemize}

\subsection{OF Statistics Gathering}
SDN-enabled NFEs have counters \cite{of_1_3_5} which can count different port characteristics such as sent and received packet and number of received packet with errors per port. This information is retrieved on either a per NFE-basis or port-basis by the SDN controller by sending an OF \textit{Port Stats Request} message to the relevant NFE with the port number specified if retrieving information on a per-port basis.

To retrieve the following characteristic of a link:
\begin{itemize}
\item \textbf{Utilization}---is calculated by first obtaining the number of transmitted bytes $A$ being sent over a links by sending to the source NFE of the link an OF \textit{Port Stats Request} message specifying the port number of the source port of the link so that only statistics about this port is retrieved and not for all the ports of the NFE \cite{of_1_3_5}. The utilization $U$ of a link can be calcutated as:
\begin{equation}
\label{eq:util}
U=\frac{A[t]-A[t-1]}{\delta b}
\end{equation}
where $\delta$ is the time period between the 2 times at which the number of transmitted bytes $A$ is polled from the NFE. $\delta$ can be adjusted according to the accuracy and overhead required, a smaller $\delta$ will lead to a more accurate $U$ but more overhead in terms of processing and bandwidth used by the control messages.
\item \textbf{Quality}---(based on packet loss and frame errors) is calculated by first obtaining the statistics of the source ($s,i$) and destination ($d,i$) ports of the link in the same manner as above for utilization. The link quality $Q$ can be calculated by using:
\begin{align}
Q=& (A_{s,i}-A_{s,i}^{\text{Dropped}}-A_{s,i}^{\text{Error}}-R_{d,i})+(R_{d,i}^{\text{Dropped}}+\nonumber \\
&R_{d,i}^{\text{Error}})+(A_{s,i}^{\text{Dropped}}+A_{s,i}^{\text{Error}})\nonumber \\ \label{eq:pkt_loss}
\end{align}
where $A_{s,i}$, $A_{s,i}^{\text{Dropped}}$ and $A_{s,i}^{\text{Error}}$ are respectively the total no. of packets transmitted, dropped and containing errors on the transmission pipeline of port $i$ of source NFE $s$. $R_{d,i}$, $R_{d,i}^{\text{Dropped}}$ and $R_{d,i}^{\text{Error}}$ are respectively the total no. of packets received, dropped and containing errors on the reception pipeline of port $i$ of destination NFE $d$. 
\end{itemize}

\subsection{Optimizing the Network Monitoring Process}
The network monitoring process can be optimized through the following:
\begin{itemize}
\item the monitoring information is shared between the different RNNs by re-using the monitoring information already in the monitoring databases if it is recent. This is different to traditional CPN \cite{gelenbe2002} where each NFE will send smart packets for each of its RNNs to collect network state. This allows CRE to have a worst-case/upper bound on its network monitoring activity  that is the same as when the current global SDN traffic engineering applications \cite{akyildiz2014} poll every NFEs in the network. Moreover, CRE runs a global and optimal routing algorithm when it finds itself in a scenario where it is monitoring the whole network.
\item the NMM will not repeat the network monitoring if last monitoring was done less than threshold time $t_e$ ago. $t_e$ can be seen as a trade-off between having updated network state and overhead associated with network monitoring. In addition, $t_e$ is an adjustable parameter that can be optimized for e.g., depending on the rate of change of the link characteristic. If the link characteristic is stable, then a longer $t_e$ can be used \cite{van2014}.
\item OF statistics are stored in the database and reused if appropriate and needed. For e.g., 1 RNN may request the NMM to provide the quality of a link and shortly after, another RNN may request for the utilization of the same link. Since OF returns all the port counters after an OF \textit{Port Stats Request} message \cite{of_1_3_5}, the NMM may have the updated information already in its database without having to poll the involved NFE(s) again.
\end{itemize}

\section{PTM: Path-to-OF Translator Module}
\label{lb:PTM}
The main objective of the Path-to-OF Translator Module (PTM) is to install and/or update OF flow rules in the network so that the traffic flow follows the path chosen by CRAM without any packet loss, forwarding loops and unecessary \textit{Packet-In} messages being sent to the SDN controller.
PTM operates in the following manner if the path is made of 2 or more NFEs:
\begin{itemize}
\item New path---OF rules are inserted at the NFEs by starting with the last NFE of the path to the first NFE so that \textit{Packet-In} are not triggered by the NFEs for flows for which CRAM has already calculated a path. For the last NFE, a rule is added which matches the original flow with the addition of VLAN ID 1. The actions of the rule is to remove the VLAN tag and output the packet to the destination port. Next, the rule with the same match as the previous rule is added to all intermediate NFEs except the first NFEs of the path. The action of the rule is to output to the relevant port as selected by CRAM. Finally, for the first NFE of the path, the rule is to match the orginal flow only with actions being tag the flow packets with VLAN ID 1 with action of forwarding the flow to the output port connected to the next NFE of the path.
\item Change existing path---OF rules are inserted for all NFEs except the first NFE as the previous case with the difference that the VLAN ID used is 2 if 1 was previously used for the flow and 1 if 2 was used. The use of VLAN tags is used to avoid conflict between the old and new rules for the same flow. The OF rule for the flow in the first switch is modified so that now it tags the flow packets with the new VLAN ID and outputs the packet to the port connected to the next NFE of the new path. Finally, the old rules in the intermediate and last NFE of the old path are deleted.   
\end{itemize}

It should be noted that if the source and destination of the flow is directly connected to the same NFE, CRE does not need to calculate a path, it just inserts a rule which matches the flow with the single action of outputing the flow packets to the port connected to the destination NFE.
 
In order to automatically remove rules for flows with no traffic for the last $t_{i}$ seconds, the \textit{idle\_timeout} is set to $t_{i}$ for all the rules inserted in the NFE. The \textit{hard\_timeout} is set to 0 so that the network always has a path for a current active flow even if the controller becomes disconnected.

\section{Evaluation}
The following evaluations scenarios where carried out using the SDN emulator Mininet \cite{Lantz_2010} where custom topology can be deployed with custom delay, bandwidth, packet loss and queue length for each link.

\subsection{Scenario 1: Delay Detection and Path Switching in illustrative network}
The aim of scenario 1 is to demonstrate CRE finding, monitoring and switching paths in a network where the delay on links can vary. Scenario 1 uses the topology in Fig.~\ref{fg:net_arch} where the initial link delays are shown in the figure. When Ping \#1 enters the network with source $1$ and destination $2$, CRE first installs 2 paths $G_{1,2}=H_1 \rightarrow N_1 \rightarrow N_2 \rightarrow H_2$ and $G_{2,1}=H_2 \rightarrow N_2 \rightarrow N_1 \rightarrow H_1$ based on shortest path based on hop count which also coincidentally gives the paths with the shortest delay with the Ping Round-Trip Time (RTT) of 120$ms$, the higher RTT for Ping \#1 compared to subsequent Pings are due to the overhead of notifying the controller of new flows, calculating paths and inserting rules inside the NFEs. Then CRE starts to monitor the network based on RNN with RL. At time 9$s$ in the experiment, the links from/to $N_1$ to/from $N_2$ $L_{1,2}$ and $L_{2,1}$ suffers from increased delay from 25$ms$ to 200$ms$ resulting in the Ping RTT to increase to 420$ms$. CRE detects this increase in path delays through its monitoring of the network, the first path that is changed is $G_{2,1}$ where an alternative path $G_{2,1}=H_2 \rightarrow N_2 \rightarrow N3 \rightarrow N_1 \rightarrow H_1$ is found which reduces the Ping RTT to 
280$ms$ at Ping \#18. Furthermore, the path $G_{1,2}$ is changed to the $G_{1,2}=H_1 \rightarrow N_1 \rightarrow N3 \rightarrow N_2 \rightarrow H_2$ with the RTT becoming 140$ms$ at Ping \#20. These two new paths found by CRE are the optimal one for the new network conditions.

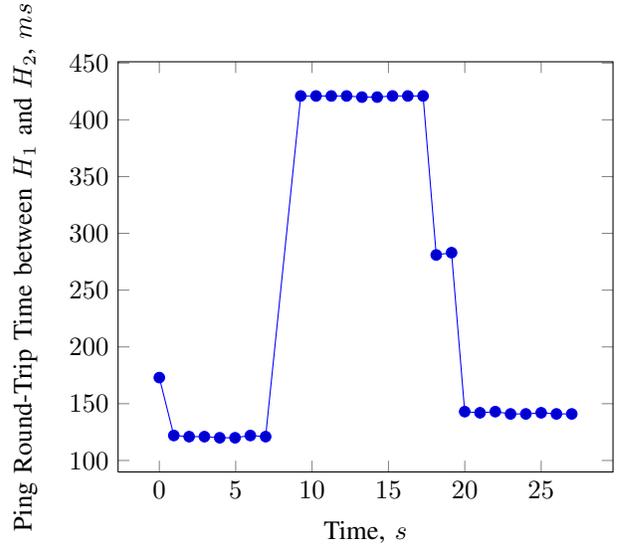
\begin{figure}
\begin{tikzpicture}
	\begin{axis}[
		width=0.45\textwidth,
		xlabel={Time, $s$},	
		ylabel={Ping Round-Trip Time between $H_1$ and $H_2$, $ms$},
		ytick        = {100,150,...,450},
	]

\addplot coordinates {
(0,173)
(0.9498109818,122)
(1.9509010315,121)
(2.9525299072,121)
(3.9532279968,120)
(4.9562759399,120)
(5.9587910175,122)
(6.9598698616,121)
(9.2590529919,421)
(10.2604658604,421)
(11.2610108852,421)
(12.2620780468,421)
(13.2628979683,420)
(14.2647688389,420)
(15.2658128738,421)
(16.2671978474,421)
(17.2688388824,421)
(18.1297729015,281)
(19.1320369244,283)
(19.9942269325,143)
(20.9945569038,142)
(21.9964938164,143)
(22.9956169128,141)
(23.9974968433,141)
(24.9993879795,142)
(25.9996960163,141)
(27.0013699532,141)
};

\end{axis}
\end{tikzpicture}
\caption{Change in Ping Round Trip Time when the delay on links $L_{1,2}$ and $L_{2,1}$ is changed to 200ms in topology of Fig. 1} 
\label{fig:topo1_delay_exp}
\end{figure}

\subsection{Scenario 2: CRE convergence and monitoring reduction in the GEANT operational network}
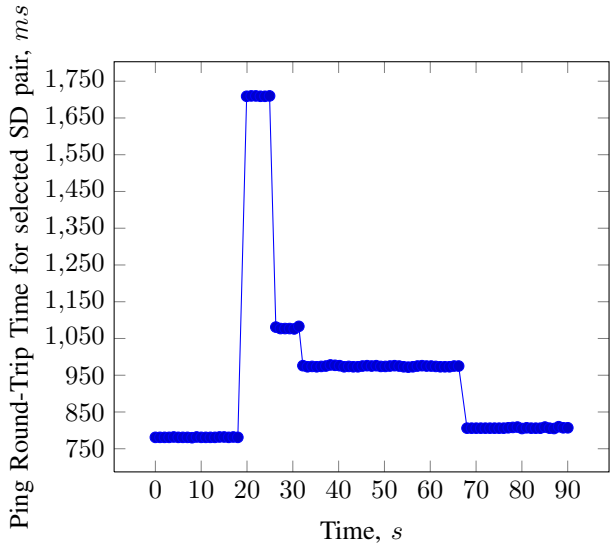
\begin{figure}
\begin{tikzpicture}
	\begin{axis}[
		width=0.45\textwidth,
		xlabel={Time, $s$},	
		ylabel={Ping Round-Trip Time for selected SD pair, $ms$},
		ytick        = {750,850,...,1800},
		xtick		={0,10,...,90}
	]

\addplot coordinates {
(0,781)
(0.999448061,781)
(1.9989681244,781)
(2.9980239868,781)
(3.9986691475,782)
(4.9977359772,781)
(5.9980070591,781)
(6.9976751804,781)
(7.9974200726,780)
(8.9982759953,782)
(9.9978041649,781)
(10.9973092079,781)
(11.9968931675,781)
(12.9971780777,781)
(13.9971671104,782)
(14.9975690842,782)
(15.9968321323,781)
(16.9977490902,782)
(17.9987661839,781)
(19.9241931438,1709)
(20.924336195,1710)
(21.9239270687,1710)
(22.9233381748,1709)
(23.9231071472,1709)
(24.9243731499,1710)
(26.2947621346,1081)
(27.2976222038,1077)
(28.2992441654,1077)
(29.3013210297,1077)
(30.3020470142,1076)
(31.3092081547,1083)
(32.2045371532,976)
(33.2009880543,973)
(34.203286171,974)
(35.2030770779,973)
(36.2035200596,974)
(37.2040140629,975)
(38.2072169781,978)
(39.2067980766,977)
(40.204788208,976)
(41.2023839951,973)
(42.2028911114,974)
(43.2021520138,973)
(44.2007510662,973)
(45.2029972076,975)
(46.2030692101,976)
(47.2016210556,975)
(48.2020761967,976)
(49.1996841431,974)
(50.1995279789,974)
(51.1996920109,975)
(52.2001821995,976)
(53.1988270283,975)
(54.1965429783,973)
(55.1956560612,972)
(56.1958701611,973)
(57.1960530281,975)
(58.1962530613,976)
(59.1953520775,975)
(60.1955401897,975)
(61.1942811012,974)
(62.193696022,973)
(63.1922671795,973)
(64.1921901703,973)
(65.1937720776,975)
(66.1938500404,975)
(68.0255072117,806)
(69.0244569778,806)
(70.0245420933,806)
(71.0239880085,806)
(72.0240521431,806)
(73.0243730545,806)
(74.0239369869,806)
(75.0241510868,806)
(76.0237231255,806)
(77.0245041847,807)
(78.02541399,808)
(79.0263371468,809)
(80.0227351189,805)
(81.023733139,807)
(82.0229620934,806)
(83.0232071877,806)
(84.0218760967,806)
(85.0257840157,809)
(86.021490097,806)
(87.0211532116,805)
(88.0254361629,810)
(89.0213739872,807)
(90.0215651989,807)

};

\end{axis}
\end{tikzpicture}
\caption{Change in Ping Round Trip Time for a given Source-Destination pair when the network is loaded around 20$s$ in the experiment and CRE is active.} 
\label{fig:geant_delay_exp}
\end{figure}

The network topology used is GEANT---a European academic network which is made up of 23 Point of Presence (PoP), represented as NFEs in this experiment, and 74 unidirectional links, where the propagation delay of given link is calculated using the line-of-sight distance between the source and destination PoP of the link. The experiment objective is to measure the amount of time and monitoring probes that CRE takes to discover the best path in the network for a given Source-Destination (SD) pair compared to an optimal routing algorithm which probes all the 74 links each time that it probes the network. The experiments are run for 90$s$ with the network being loaded around 20$s$ into the experiment. The monitoring frequency of both CRE and the optimal algorithm is set to 5$s$ Fig.~\ref{fig:geant_delay_exp} shows the increase in Ping RTT around 20$s$ when the network is loaded and the decrease in RTT when CRE finds iteratively better paths to route the flow.

Table.~\ref{tab:geant_results} shows that CRE can reduce by up to 9.5 times the amount of monitoring required to find a best path which is on average only 1.65\% worse than the optimal RTT and takes around 30.8$s$ (with a 5$s$ minimum update frequency) more time to find. 

\begin{table}\centering
\caption{CRE convergence vs Optimal}
\begin{tabular}{|c|c|c|c|c|}
  \hline
  Exp. No. & No. of & No. of  & CRE addit-  & Increase in\\
   & Optimal & CRE  & ional time  &  CRE Ping RTT\\
  & Monitoring &  Monitoring & over Optimal & over Optimal\\
  & & & ($s$) & (\%) \\
  \hline
  1	& 370	& 23	& 32.6 &  3.80\\
  2	& 370	& 69	& 39.6	    &  0 \\
  3	& 296	& 43	& 44.4 &  2.03\\
  4	& 370	& 21	& 1.01		&  0\\
  5	& 370	& 31	& 36.5 &  2.41\\
  \hline
  \hline				
  Avg.&	355.2 	&37.4	&30.8	&  1.65\\
  \hline
\end{tabular}\label{tab:geant_results}
\end{table}

\section{Conclusions and Future Work}
In this work, we show that CRE is able to find paths which are very close to the optimal ones without incurring a large monitoring overheard. CRE is able to achieve this performance without requiring any modification to already-deployed SDNs which run OpenFlow. 

It is our intention to further improve the evaluation section of this work by carrying more sophisticated experiments with different CRE parameters, network topologies, traffic conditions, QoS metrics and network events such as link and node failures.

\bibliographystyle{IEEEtran}

\bibliography{IEEEabrv,sdn_ref_icl}

\end{document}